\documentclass[useAMS,usenatbib]{mn2e}
\usepackage{graphicx}

\newcommand{\lsun}{L$_{\odot}$}
\newcommand{\msun}{M$_{\odot}$}
\newcommand{\kms}{km s$^{-1}$}

\title[Clumps along the CB3 outflow]{A detailed modelling of the chemically rich clumps along the CB3 outflow}
\author[Benedettini et al.]{Milena Benedettini$^{1,2}$\thanks{E-mail:
milena.benedettini@ifsi-roma.inaf.it}, Jeremy A. Yates$^{2}$, Serena Viti$^{2}$ and Claudio Codella$^{3}$
\\
$^{1}$INAF-Istituto di Fisica dello Spazio Interplanetario, Area di Ricerca di Tor 
Vergata, via del Fosso del Cavaliere 100, 00133 Roma, Italy \\
$^{2}$Department of Physics and Astronomy, University College London, Gower Street, London, WC1E6BT, UK\\
$^3$INAF-Istituto di Radioastronomia, Sezione di Firenze, Largo E. Fermi 5, 
50125 Firenze, Italy}
\begin{document}

\date{Accepted . Received .}

\pagerange{\pageref{firstpage}--\pageref{lastpage}} \pubyear{2005}

\maketitle

\label{firstpage}

\begin{abstract}
In order to investigate the origin and the structure of the low
velocity, chemically rich clumps observed along the lobes of low- and
intermediate-mass outflows, we construct a detailed model of the S1
clump along the CB3 outflow. We use a time-dependent chemical model
coupled with a radiative transfer model to reproduce the observed line
profile for a direct comparison with previous observations of this
clump. We find that the simultaneous fitting of multiple species and
transitions is a powerful tool in constraining the physical parameters
of the gas. Different scenarios for the clump formation have been
investigated. The models that better reproduce all the observed lines
are those where the clump is formed, at least partially, before the
advent of the outflow; with the advent of the outflow the clump
undergoes a short period of non-dissociative shock and the consequent
release of the icy mantle together with the high temperature chemistry
leads to the observed chemical enrichment. Our results also suggest
the presence of substructure within the clump: a more extended
component traced by CS, SO and the lower energy transitions (3$_{\rm
K}$-2$_{\rm K}$ and 2$_{\rm K}$-1$_{\rm K}$) of CH$_3$OH, and a more
compact component traced by H$_2$CO, SO$_2$ and the higher energy transitions
(5$_{\rm K}$-4$_{\rm K}$) of CH$_3$OH.
\end{abstract}


\section{Introduction}

Bipolar molecular outflows are ubiquitously present around Young
Stellar Objects (YSOs) and are associated with the very early stages
of star formation (e.g. \citealt{richer00}).

A large range of excitation conditions and a significant chemical
differentiation are present along the lobes of the outflows. The
shock--excited gas cools mainly radiatively via the emission of atomic
and molecular species whose abundances are strongly enhanced by the
shocks generated by the interaction between the protostellar wind and
the molecular cloud. High velocity (v$\ge$100 km s$^{-1}$) clumps, the
so called molecular bullets, are detected along the axis of some
outflows. The prototype of such bullets are those observed in SiO in
the lobes of the L1148 outflow \citep{bachiller1991} but they
generally show weak emission of other molecular lines. A few outflows
however, e.g. L1157 and CB3 (\citealt{bachiller2001},
\citealt{codella99}), have a particular rich emission spectrum with
molecular clumps at relatively low velocity (v$\le$ 10 km
s$^{-1}$). These molecular clumps show emission from molecular species
usually unobserved along outflows and they have dimensions which do
not exceed 0.1 pc. While the high velocities and high abundance of SiO
suggest that the molecular bullets are most likely associated with
mini-bow shocks formed by the outflow propagation
(e.g. \citealt{deutrey97}), the origin of the low velocity clumps is
not yet clear. In a recent study \citet{viti04} (thereafter Paper I),
have investigated two possibilities for the origin of the low velocity
clumps: i) the clumps are formed by episodic mass loss of the forming
object or ii) the clumps are density structure pre-existing the advent
of the outflow and the effect of the outflow is to accelerate and shock them,
altering their chemistry.  Paper I used a chemical model to simulate
the clump formation and its subsequent interaction with the outflow
and inferred that the low velocity outflow clumps are probably density structures
formed, at least partly, prior of the advent of the outflow. With the
advent of the outflow, not only the temperature of the clumps
increases ($\sim$ 100 K), but the clumps also undergo a short period
of non-dissociative shock. This preliminary conclusion was reached by
comparing the column densities derived by the chemical models with
existing single dish observations along the CB3 \citep{codella99} and
L1157 outflows \citep{bachiller2001}.
The shortcomings of such an analysis were that the estimations of the
column densities derived from the chemical models, as well as those
derived from the observations, suffer from high uncertainties due to the
fact that some of the parameters required to calculate them may be unknown. In
particular, to estimate the observed column densities arbitrary
assumptions have to be made regarding LTE conditions, excitation
temperatures and the lines being optically thin. On the other hand,
the theoretical calculation of column densities from the chemical
model required knowledge of the geometry of the emitting
region. However, when dealing with small emitting regions, such as the
low velocity clumps along molecular outflows, the observations
(usually single dish with spatial resolution of few tens of arcsec) do
not constrain such geometry. If more transitions of the same species
are available, one can use a Large Velocity Gradient (LVG) model to
derive the density and temperature of the emitting gas but even in
such cases often the results have a large uncertainty because the
observed line ratios are not always sensitive to the temperature or
density.
 
One way to reduce the number of arbitrary assumptions is to model {\it
directly} the observed line emission using a radiative transfer
model that predicts the line profile, a method already adopted by
other authors (e.g. \citealt{jorge04}, \citealt{doty04} and
\citealt{schoier02}). In this work we attempt such an approach by
coupling the time dependent chemical model of Paper I with the
radiative transfer model Spherical Multi-Mol (SMMOL)
\citep{rawlings01}. In particular, we expand the grids of chemical
models computed in Paper I and use them as input parameters for the
radiative transfer model to simulate the line profiles of several
species and, where possible, of multiple transitions of the same
species. We then simultaneously fit such profiles with the
observations of CB3 and attempt to constrain the physical parameters
of the emitting gas and thus their formation scenario. The paper is
organized as follow: in Section 2 both the chemical and the radiative
transfer models are briefly described. In Section 3 the models are
compared with the lines detected towards a clump in the CB3
outflow. The results are presented in Section 4 and discussed in Section 5.

\begin{figure}
\includegraphics[width=8cm]{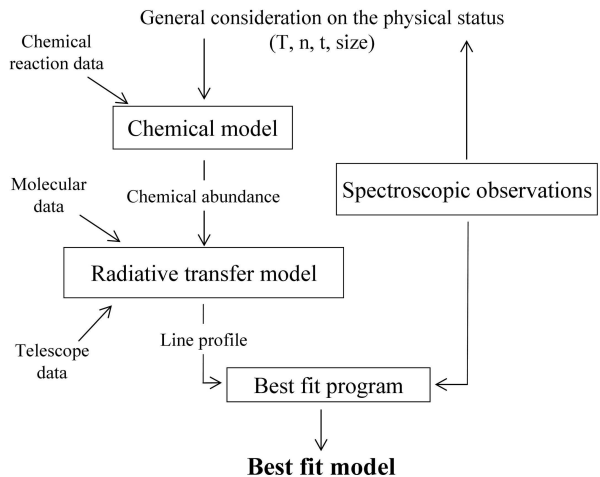}
\caption[]{A flow diagram of the scheme applied to compare the
observed data to the theoretical models in order to derive the best
fit model.}
\label{scheme}
\end{figure}

\section{The chemical and radiative transfer models}

A flow diagram of the method used to compare the observed data with
the theoretical models is shown in Fig. \ref{scheme}. A first order
estimate of the general physical conditions of the clump, such as the
size, the density and the temperature, is derived from the
observations and used to describe the physics of the phenomenon in the
chemical model. The chemical model predicts the fractional abundances;
these are given as input to the radiative transfer model together with
the density, the temperature and the beam of the telescope to produce
the line profiles that are $directly$ compared with the observed
profiles.  Recently, the same method has been used for modelling the
envelope of low mass protostars (\citealt{jorge04};
\citeyear{jorge05}; \citealt{schoier02}; \citealt{doty04}) and it was
found to be a powerful tool to probe the physical and chemical
structure of the protostellar envelopes as well as the age of the
protostar.

The time-dependent chemical model is described in detail in Paper
I. Here we summarize the important characteristics: the chemical
network is taken from the UMIST database and includes 221 species
involved in 3194 gas-phase and grain reactions. The model is a
two-phase calculation. Phase I simulates the formation of a
pre-existing clump, or of the homogeneous dark cloud, depending on
what scenario we are modelling, and includes freeze-out of 
the gas on to the grains (see Paper I and below). Phase II
simulates the advent of the outflow by warming the clump to 100-200 K
or, in the presence of a non-dissociative shock, by increasing the gas
temperature to 1000 K for 100 yr and then cooling it down to 100-200
K. As a reminder, the four scenarios considered here (as in Paper I) are:

\begin{itemize}
\item Grid A: a pre-existent clump at uniform density warmed by the outflow. In this paper we introduce new models, called As, where the clump at uniform single density is shocked by the outflow;
\item Grid B: a pre-existing clump with a density structure shocked by the outflow;
\item Grid C: a pre-existing clump with a density structure warmed by the outflow;
\item Grid D: a clump formed and warmed or shocked by the outflow. 
\end{itemize}

For each grid different conditions are investigated, changing the
model parameters: the depletion efficiency, the type of collapse (free
fall or retarded), the grain mantle evaporation (instantaneous or
time-dependent), the final density and temperature, the initial
sulphur abundance (see Section 4.1), the fraction of frozen H$_2$CO
and CO converted into CH$_3$OH (see Section 4.4), for a total of 33
models. Each model is time-dependent and evolves for 10$^5$ yr. See
Paper I for a full description of the model and its assumptions, as
well as of each scenario.

The chemical abundances produced by the chemical model together with
the density and temperature profile are used as input for the
radiative transfer model SMMOL \citep{rawlings01}, an approximate
$\Lambda$-iteration (ALI) code that solves multi-level non-LTE
radiative transfer problems. The molecular data needed for the
calculation are taken from the
LAMBDA\footnote{http://www.strw.leidenuniv.nl/$\sim$moldata} public
database \citep{schoier05}, except the CH$_3$OH-A data which we
obtained from D. Flower (private communication).  As first step the
code calculates the total radiation field and the level populations
assuming LTE and the interstellar radiation field as input continuum
\citep{black1994}. It then re-calculates the total radiation field,
checks for convergence and repeats the process until convergence is
achieved. At each radial point the code generates the level
populations and the radiation field. The emergent intensity
distributions are then convolved with the telescope beam, so that the
model directly predicts the line profiles for a given source as
observed with a given telescope (e.g. the IRAM 30-m telescope in this
work). The radiative transfer model has been successfully benchmarked
with other radiative transfer codes \citep{vanza02}. For each model
the radiative transfer program has been run for a selected subset of
the time steps of the chemical model for a total of 19 time steps
starting from 10$^3$ yr after the formation of the clump to 10$^5$ yr.

\section{Modelling the clumps along CB3}

\begin{table}
\caption{List of the molecules and transitions observed in the S1 clump of the CB3 outflow that are modeled with the radiative transfer code.}
\label{specie}
\begin{tabular}{|l|c|r|}
specie & line & $\nu$ (GHz) \\
\hline
\hline
CO & 2-1 & 230.538\\
CS & 3-2 & 146.967\\
SO & 6$_5-5_4$ & 219.949\\
SO & 4$_3-3_2$ & 138.178\\
SO$_2$ & 3$_{13}$-2$_{02}$ & 104.029\\
H$_2$CO  & 3$_{21}$-2$_{20}$ & 218.760\\
CH$_3$OH & 5$_0$-4$_0$ A$^+$ & 241.791\\
CH$_3$OH & 5$_{-1}$-4$_{-1}$ E & 241.767\\
CH$_3$OH & 5$_0$-4$_0$ E & 241.700\\
CH$_3$OH & 3$_0$-2$_0$ A$^+$ & 145.103\\
CH$_3$OH & 3$_{-1}$-2$_{-1}$ E & 145.097\\
CH$_3$OH & 3$_0$-2$_0$ E & 145.094\\
CH$_3$OH & 2$_0$-1$_0$A$^+$ & 96.741\\
CH$_3$OH & 2$_{-1}$-1$_{-1}$ E & 96.739\\
\end{tabular}
\end{table}

CB3 is a Bok globule at a distance of 2500 pc associated with star
formation. In particular, it hosts an intermediate-mass YSO driving a
4 \msun\, outflow with a mechanical luminosity of 5.6 \lsun. The
outflow has been mapped with the IRAM 30-m telescope in a number of
species \citep{codella99} and four molecular clumps were identified
along the main axis. The physical conditions derived in the four
clumps by \citet{codella99} by means of LVG or LTE calculations are
very similar to each other, and we use these values to guide the
choice of the input parameters in our chemical models. We choose to
fit the clump S1 of the CB3 outflow located in the southern lobe at an
offset (0\arcsec,-30\arcsec) from the millimeter driving source since it is the best defined clump. In Table \ref{specie} we list all the species and the
lines observed in the S1 position for which the molecular data
(Einstein coefficients and collisional rates) exist and the radiative
transfer code has been run. Most of these lines have a Gaussian
profile with similar width.
We adopt a microturbulent velocity of 5 km s$^{-1}$ for our radiative transfer modelling. This is equivalent to modelling the observed line profiles with a single gaussian line with a fixed FWHM of 10 km s$^{-1}$. We assume a clump size of 0.12 pc as derived from the IRAM maps. However, we note that most of the emission from the clump is unresolved by the large IRAM-30m beam (HPBW = 10\arcsec -- 29\arcsec), so 0.12 pc is most likely an overestimate of the real size of the clump. In Table \ref{model} we list the
chemical models computed and their parameters: the name of the model
(column 1); the presence (Y) or absence (N) of non--dissociative shock
(column 2); the density of the clump (column 3), for models with a
density profile we give the two extremes of the range; the final
temperature (column 4); the percentage of gas depleted into grains at
the end of the Phase I (column 5); the initial sulfur abundance
(column 6); and any other relevant parameters that were varied (column
7) such as the type of collapse simulated in Phase I (free-fall or
retarded), the percentage of CO or H$_2$CO converted into CH$_3$OH on
the grain surfaces or the type of evaporation from the grains once the
outflow approaches the clumps (instantaneous or time-dependent as in
\citet{viti99}). For the models of Grid D we also give the velocity of
the outflow. For this work, we extend the grids of models already
published in Paper I.
In particular, we computed new models for: clumps at uniform density
that are shocked (Grid As); clumps with a density profile from
5$\times$10$^5$ cm$^{-3}$ to 10$^6$ cm$^{-3}$ at a temperature of
T=200 K; clumps with a density profile from 2.5$\times$10$^5$
cm$^{-3}$ to 5$\times$10$^5$ cm$^{-3}$ and from 3$\times$10$^5$
cm$^{-3}$ to 6$\times$10$^5$ cm$^{-3}$. From Paper I, we found that
most of the models from Grid D, where the clumps are formed by the
compression of the incoming outflow, give unphysical results and thus
in this paper we restrict our analysis to models D1, D3, D4, D4-shock
and D8 only.
\begin{table*}
\caption{List of the models and their parameters: presence (Y) or absence(N) of non--dissociative shock, density, final temperature, percentage of freeze-out, initial sulfur abundance and other possible parameters that are different with respect to the other models. For Grid D models the density and temperature are the final ones (at t=10$^5$ yr).}
\label{model}
\begin{tabular}{|l|c|c|c|c|c|c|}
model & shock & n(H$_2$) & T & FR & X(S) & Note \\
      &       &(10$^5$ cm$^{-3}$) & (K)& (\%) & (10$^{-7}$) &  \\
\hline
\hline
A1 &  N    & 1  & 210 & 15 &  130 & \\
A2 &  N    & 1  & 210 & 35 &  130 & retarded collapse\\
A3 &  N    & 1  & 210 & 11 &  130 & retarded collapse\\
A4 &  N    & 1  & 210 & 20 &  19  & \\
A5 &  N    & 10 & 210 & 20 &  130 & \\
A6 &  N    & 10 & 210 & 30 &  1.3 & \\
A7 &  N    & 10 & 210 & 55 &  1.3 & \\
A8 &  N    & 10 & 210 & 80 &  1.3 & \\
A9 &  N    & 50 & 210 & 25 &  130 & \\
A10 & N    & 5  & 210 & 30 &  1.3 & \\
\hline
As1 & Y    & 5  & 210 & 30 &  1.3 & \\
As2 & Y    & 1  & 210 & 30 &  1.3 &\\
\hline
B1 &  Y   & 5-10& 110 & 40 &  1.3 &\\
B2 &  Y   & 5-10& 110 & 60 &  1.3 &\\
B3 &  Y   & 5-10& 110 & 60 &  1.3 & 100\%H$_2$CO$\Rightarrow$CH$_3$OH\\
B4 &  Y   & 5-10& 110 & 80 &  1.3 & 100\%H$_2$CO$\Rightarrow$CH$_3$OH\\
B5 &  Y   & 5-10& 110 & 60 &  1.3 & 10\%CO$\Rightarrow$CH$_3$OH\\
B6 &  Y   & 5-10& 110 & 60 &  1.3 & 20\%CO$\Rightarrow$CH$_3$OH\\
B7 &  Y   & 5-10& 210 & 60 &  1.3 &\\
B8 &  Y   & 2.5-5& 210 & 40 & 1.3 &\\
B9 &  Y   & 3-6  & 210 & 60 & 1.3 &\\
B10 & Y   & 5-10& 210 & 20 &  1.3 &\\
B11 & Y   & 3-6 & 210 & 40 &  1.3 &\\
B12 & Y   & 2.5-5& 210 & 60 & 1.3 &\\
\hline
C1 &  N   & 5-10& 110 & 40 &  1.3 &\\
C2 &  N   & 5-10& 110 & 60 &  1.3 &\\
C3 &  N   & 5-10& 110 & 60 &  1.3 & 100\%H$_2$CO$\Rightarrow$CH$_3$OH, time-dependent evaporation\\
C4 &  N   & 5-10& 210 & 60 &  1.3 &\\
\hline
D1 &  N   & 5-10& 110 & 20 &  1.3 & v$_{outflow}$= 2 km s$^{-1}$\\
D3 &  N   & 5-10& 110 & 20 &  1.3 & v$_{outflow}$= 100 km s$^{-1}$\\
D4 &  N   & 2.5-5&100 & 20 &  1.3 & v$_{outflow}$= 2 km s$^{-1}$\\
D4-shock& Y & 5-10&100& 20 &  1.3 & v$_{outflow}$= 2 km s$^{-1}$ \\
D8 &  N   & 5-10& 100 & 20 &  1.3 & v$_{outflow}$= 20 km s$^{-1}$\\
\end{tabular}
\end{table*}

To fit the lines listed in Table 1 we used a $\chi ^2$ method that
simultaneously fits the total integrated flux of the observed
transitions of various species. The formula used is:

\begin{equation}
\chi^2=\frac{1}{N}\sum_{i=1}^{N}\left[\frac{Flux_{mod}(i)-Flux_{obs}(i)}{Flux_{obs}(i)}\right]^2
\end{equation}

\noindent where $i$ is the index indicating the line, $N$ is the total number of lines considered, $Flux_{mod}$ is the integrated flux of the model and $Flux_{obs}$ is the observed integrated flux. The difference between the theoretical and observed flux is weighted with the observed flux instead of the error of the measurement because the uncertainties associated to the chemical models are higher than the error of the observations, making it meaningless to use the error as the weight. Moreover, in this way we avoid giving the brighter lines a higher weight. A similar formula has also been used by other authors, e.g. \citet{doty04}. 

\section{Results}
In the next 5 subsections, we will analyse the results we obtain for
the fit of each molecular species separately; in Section 5 we will
attempt to fit all the species simultaneously. In order to take into
account the uncertainties in the parameters adopted in the models and
the calibration errors of the observations, we assume that we obtain a
good match to the models when the theoretical line flux differs less
than 10\% from the observed flux.

\subsection{SO}

As already noted in Paper I, if in the chemical model we assume a
standard solar initial abundance of sulfur (1.3$\times$10$^{-5}$) the
final abundance of SO is much higher than the observed
one. Moreover the radiative transfer model produces strong self-absorption in
the lines, which is not present in the observed line profile. The
observed SO lines can be better reproduced assuming a depleted sulfur
initial abundance of 1.3$\times$10$^{-7}$ as also indicated by
previous studies (\citealt{viti03}; \citealt{ruffle99}).

\begin{figure}
\includegraphics[width=3.5cm,angle=90]{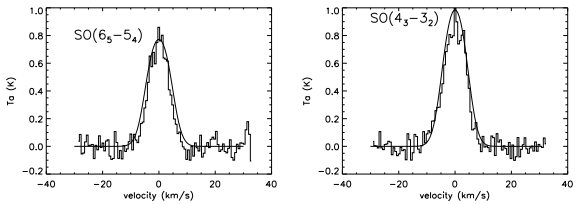}
\caption{The two SO 6$_5-5_4$ and 4$_3-3_2$ lines observed in the clump S1 of the CB3 outflow (histogram) with the line profiles predicted by the best fit model B10 at t=1$\times10^3$ yr (continuum line).}
\label{sofitfig}
\end{figure}

We have detected two lines of SO and, due to the different excitation
temperature of these lines (24 and 11 K), their ratio should allow us
to constrain the temperature of the gas. Indeed, we see that models
with temperature of T$\sim$200 K are the only models that can
reproduce the observed line ratio. In Table \ref{sofit} we list all
the models where both the two SO line fluxes differ $\le$10\% from the
observed fluxes, ordered by increasing $\chi^2$. The SO column density
in all the models ranges from 3.6$\times$10$^{14}$cm$^{-2}$ to
6.4$\times$10$^{14}$cm$^{-2}$. The models from Grid B are the best
ones, while models without a non-dissociative shock and models with
low density (n(H$_2$)$<$3$\times$10$^5$ cm$^{-3}$) do not seem to fit
the SO lines at all. In Table \ref{sofit} we
note that models with equal input parameters but two different
percentage of freeze-out (B10 and B7 or B9 and B11) can both fit the
observed lines but at different epoch. In particular, the models with
the lower percentage of freeze-out fit the lines at an early epoch
with respect to the models with higher freeze-out. The best fit is
given by the model B10 (n(H$_2$)=(5-10)$\times10^5$ cm$^{-3}$,
T$\sim$200 K, FR=20\%, Shock=Y) at a time t=10$^3$ yr with
N(SO)$\sim$4$\times$10$^{14}$cm$^{-2}$. However, in the B10 model the
SO column density increases with time, it has a maximum at
t=3$\times$10$^4$ yr and then it decreases again to a value similar to
the one assumed at t=10$^3$ yr. This implies that a good fit can also
be obtained at a later time of t=9$\times$10$^4$ yr.  The comparison
between the observed lines and the best fit model is shown in
Fig. \ref{sofitfig}.

\begin{table}
\caption{List of the best models for the two SO transitions, ordered by increasing $\chi^2$, with the relative epoch, gas density, temperature, freeze-out and SO column density.}
\label{sofit}
\vspace{0.3 cm}
\begin{tabular}{|c|c|c|c|c|c|}
model & epoch & n(H$_{2}$)& T & FR & N(SO) \\
      & (yr)& (10$^{5}$ cm$^{-3}$) & (K) & \% & (cm$^{-2}$)\\
\hline
B10 & 1$\times10^3$ & 5--10& 210& 20 & 3.8$\times10^{14}$\\
A10 & 5$\times10^3$ & 5    & 210& 30 & 6.4$\times10^{14}$\\
B11 & 2$\times10^3$ & 3--6 & 210& 40 & 5.5$\times10^{14}$\\
B9  & 3$\times10^4$ & 3--6 & 210& 60 & 5.3$\times10^{14}$\\
B10 & 9$\times10^4$ & 5--10& 210& 20 & 3.8$\times10^{14}$\\
B7  & 8$\times10^4$ & 5--10& 210& 60 & 3.6$\times10^{14}$\\
\hline
\end{tabular}
\end{table}

\subsection{CS}

\begin{figure}
\includegraphics[width=5cm,angle=90]{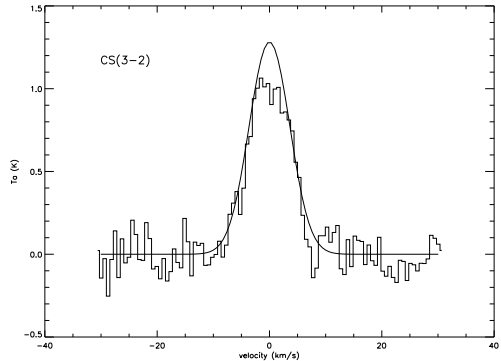}
\caption{The CS(3-2) line observed in the clump S1 of the CB3 outflow (histogram) and the line profile predicted by the best fit model As1 at t=2$\times10^3$ yr (continuum line).}
\label{csfitfig}
\end{figure}

CS is a high density tracer and we find that at a fixed density it is
also quite sensitive to the percentage of freeze-out, and in fact
models with the same temperature and density but different freeze-out
fit the line at different epochs. In 14 out of 33 models, at least a
time can be found in which the flux of the CS(3-2) line differs less
than 10\% from the observed flux (see Table \ref{csfit}) and for 3
models, namely As1, C3 and B9, we find a range of times fitting the
line. The models cover almost all the explored parameter space
indicating that the CS(3-2) line alone can not be used to constrain
the physical conditions of the clump, nor can it be used to
discriminate among different scenarios of formation. Despite the
different physical parameters, all the models in Table \ref{csfit}
have a similar CS column density: N(CS)$\sim$10$^{14}$cm$^{-2}$. It is
worth noting that the best fit model is the model As1
(n=5$\times$10$^5$ cm$^{-3}$, T$\sim$200 K, FR=30\%, Shock=Y) at a
time t=2$\times$10$^3$ yr with N(CS)=1.2$\times$10$^{14}$cm$^{-2}$. The comparison between the observed line and the best fit model is shown in
Fig. \ref{csfitfig}. The fact that the modelled line has a peak higher
than the observed one suggests that either the line has a FWHM
slightly lower than the fixed value of 10 \kms\, or the line is weakly
self-absorbed.

\begin{table}
\caption{List of the best models for the CS(3-2) transition, ordered by increasing $\chi^2$, with the relative epoch, gas density, temperature, freeze-out and CS column density.}
\label{csfit}
\vspace{0.3 cm}
\begin{tabular}{|c|c|c|c|c|c|}
model & epoch & n(H$_{2}$) & T & FR & N(CS)\\ 
      &(yr)   & (10$^{5}$cm$^{-3}$) & (K) &\% & (cm$^{-2}$)\\
\hline
As1 & 2$\times10^3$ &  5     & 210 & 30 & 1.2$\times10^{14}$\\
C1  & 1$\times10^5$ &  5--10 & 110 & 40 & 1.2$\times10^{14}$\\
A4  & 1$\times10^3$ &  1     & 170 & 20 & 1.5$\times10^{14}$\\
A2  & 3$\times10^3$ &  1     & 210 & 35 & 1.3$\times10^{14}$\\
B12 & 4$\times10^4$ &  2.5--5& 210 & 60 & 1.4$\times10^{14}$\\
B10 & 5$\times10^3$ &  5--10 & 210 & 20 & 1.6$\times10^{14}$\\
D4  & 2$\times10^3$ &  10    & 70  & 20 & 2.2$\times10^{14}$\\
As1 & 1$\times10^3$ &  5     & 210 & 30 & 1.1$\times10^{14}$\\
C3  & 6$\times10^3$ &  5--10 & 110 & 60 & 1.1$\times10^{14}$\\
D8  & 2$\times10^3$ &  10    & 70  & 20 & 1.6$\times10^{14}$\\
B9  & 4$\times10^4$ &  3--6  & 210 & 60 & 1.6$\times10^{14}$\\
C3  & 5$\times10^3$ &  5--10 & 110 & 60 & 1.2$\times10^{14}$\\
B4  & 3$\times10^3$ &  5--10 & 110 & 80 & 1.0$\times10^{14}$\\
C2  & 4$\times10^4$ &  5--10 & 110 & 60 & 1.0$\times10^{14}$\\
B9  & 3$\times10^4$ &  3--6  & 210 & 60 & 1.3$\times10^{14}$\\
As1 & 3$\times10^3$ &  5     & 210 & 30 & 1.3$\times10^{14}$\\
C3  & 5$\times10^3$ &  5--10 & 110 & 60 & 1.7$\times10^{14}$\\
B8  & 1$\times10^4$ &  2.5--5& 210 & 40 & 1.0$\times10^{14}$\\
As2 & 1$\times10^5$ &  1     & 210 & 30 & 1.2$\times10^{14}$\\
\hline
\end{tabular}
\end{table}

\subsection{SO$_2$}

In Table \ref{so2fit} we report all the models where the SO$_2$ line
flux differs $\le$10\% from the observed flux, ordered, as usual, by
increasing $\chi^2$.  The SO$_2$ column density ranges from
2$\times10^{14}$ to 5$\times10^{14}$ cm$^{-2}$. We find that N(SO$_2$)
is very sensitive to time and in all the models it increases by more
than two orders of magnitude from t=10$^3$ yr to t=10$^5$ yr. However,
we do not seem to find a good fit for times $>$ 10$^4$ yr. At later
times the modeled column density is usually higher than the observed
one. Similar to our SO fits, the models from grid B, As and A with
high H$_2$ density (n$\ge$10$^6$ cm$^{-3}$) give the best fits,
indicating once again that the scenario where the clump is (at least
partially) pre-existing the outflow is the most likely.
All models from Grid D, C and A (with the exception of A8 and A9 - see below) produce a SO$_2$ line intensity very different from the observed one, up to two orders of magnitude for the model D4. One common characteristic of these models is the lack of a post-shock phase at high temperature, hence it seems that this species is indeed a good tracer of the presence of a non--dissociative shock. This is in agreement with theoretical models of C-type shocks \citep{pineau93} where the abundance of SO$_2$ is seen to increase after the passage of the shock and SO$_2$ is the dominant form of sulfur in the post-shock region.

\begin{figure}
\includegraphics[width=5cm,angle=90]{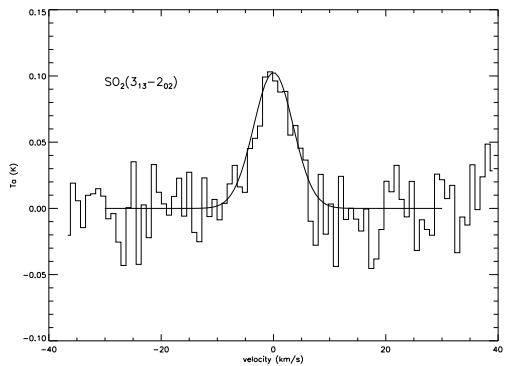}
\caption{The SO$_2$ 3$_{13}$-2$_{02}$ line observed in the clump S1 of the CB3 outflow (histogram) with the line profiles predicted by the best fit model A9 at t=6$\times10^3$ yr (continuum line).}
\label{so2fitfig}
\end{figure}

The best fit model (see Fig. \ref{so2fitfig}) is the model A9
(n=5$\times$10$^6$ cm$^{-3}$, T$\sim$200 K, FR=25\%, Shock=N) at the
epoch of t=6$\times$10$^3$ yr. We do not think, however, that this
model is correct because such high uniform density for a clump of 0.12
pc would imply a visual extinction of more than 1000 mags. Moreover,
the initial sulfur abundance for A9 is solar. Probably, in the A9
model the absence of the non-dissociative shock is compensated by the
higher initial sulfur abundance and the high density that lead to a
SO$_2$ column density able to fit the line.  As can be seen
from Table \ref{so2fit} there are quite a number of models that can
fit the SO$_2$ line. It is clear therefore that we can not use this
line to deduce uniquely the physical characteristics of our
clump. However, the SO$_2$ line does seem to trace a gas with a density
between 3$\times$10$^5$ and 10$^6$ cm$^{-3}$ and a temperature between
100 and 200 K. The freeze-out parameter is not constrained at all
since in the models that fit the data the freeze-out parameter ranges
from 25\% to 80\%. This is not surprising since SO$_2$ is a second
generation species which forms after the dissociation of other sulfur
bearing species released from the grains.

\begin{table}
\caption{List of the best models for the SO$_2$ 3$_{13}$-2$_{02}$ transition, ordered by increasing $\chi^2$, with the relative epoch, gas density, temperature, freeze-out and SO$_2$ column density.}
\label{so2fit}
\vspace{0.3 cm}
\begin{tabular}{|c|c|c|c|c|c|}
model & epoch &  n(H$_{2}$) & T & FR & N(SO$_2$) \\
     & (yr)  &(10$^{5}$ cm$^{-3}$) & (K) & \% & (cm$^{-2}$)\\
\hline
A9  & 6$\times10^3$ &  50    & 210 & 25 & 2.5$\times10^{14}$ \\
B1  & 3$\times10^3$ &  5--10 & 110 & 40 & 3.5$\times10^{14}$ \\
A8  & 2$\times10^3$ &  10    & 210 & 80 & 2.6$\times10^{14}$ \\
B8  & 5$\times10^3$ &  2.5--5& 210 & 40 & 2.7$\times10^{14}$ \\
A8  & 3$\times10^3$ &  10    & 210 & 80 & 2.7$\times10^{14}$ \\
B7  & 6$\times10^3$ &  5--10 & 210 & 60 & 5.1$\times10^{14}$ \\
As1 & 9$\times10^3$ &  5     & 210 & 30 & 2.1$\times10^{14}$ \\
B9  & 6$\times10^3$ &  3--6  & 210 & 60 & 3.3$\times10^{14}$ \\
B3  & 1$\times10^3$ &  5--10 & 110 & 60 & 3.8$\times10^{14}$ \\
A8  & 4$\times10^3$ &  10    & 210 & 80 & 2.8$\times10^{14}$ \\
B11 & 7$\times10^3$ &  3--6  & 210 & 40 & 3.5$\times10^{14}$ \\
\hline
\end{tabular}
\end{table}

\subsection{H$_2$CO}
\begin{figure}
\includegraphics[width=5cm,angle=90]{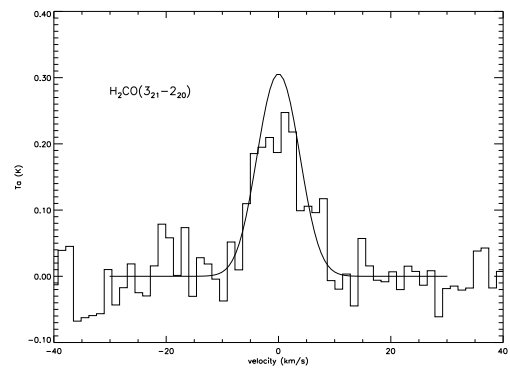}
\caption{The H$_2$CO 3$_{21}$-2$_{20}$ line observed in the clump S1 of the CB3 outflow (histogram) with the line profiles predicted by the best fit model As1 at t=1$\times10^3$ yr (continuum line).}
\label{h2cofitfig}
\end{figure}

In Table \ref{h2cofit} we report all the models that fit the observed
flux of the H$_2$CO 3$_{21}$-2$_{20}$ line within 10\%. Good fits can
be obtained in models with non-dissociative shocks (Grid B and As). In
the models without them (Grid A, C and D) the formaldehyde is usually
overabundant. This is because it is mainly formed by the reaction of
oxygen with CH$_3$. In models with a non-dissociative shock, during
the high temperature phase, the oxygen mainly interacts with neutrals
to form OH and H$_2$O, thus there is less oxygen free to form
H$_2$CO. H$_2$CO is one of the species most sensitive to time; indeed
its abundance increases by more than one order of magnitude from
t=10$^3$ yr to t=10$^5$ yr. A good fit can only be found for very
early epochs of $\le 4\times10^3$ yr. As in Paper I we find that if
the clumps evolve for more that 5$\times$10$^3$ yr after the high temperature
phase, the theoretical abundance of H$_2$CO is always overabundant with respect to the observations. Even in models where we allow conversion of H$_2$CO
to methanol on the grains (B3 and B4) an acceptable fit can only be
obtained at t=10$^3$ yr for the B3 model. The H$_2$CO column density
in the selected models ranges from 4.5$\times10^{14}$ to
6.5$\times10^{14}$ cm$^{-2}$ and the freeze-out parameter ranges
between 20\% and 60\% while the temperature is $\sim$200 K, with the
exception of the B3 model. The best fit model is As1
(n(H$_2$)=5$\times$10$^5$, T$\sim$200 K, FR=30\%, Shock = Y) at the
time t=10$^3$ yr and column density N(H$_2$CO)=6.1$\times10^{14}$
cm$^{-2}$. In Fig. \ref{h2cofitfig} the As1 model is plotted
superimposed to the observed line spectrum.

\begin{table}
\caption{List of the best models for the H$_2$CO 3$_{21}$-2$_{20}$ transition, ordered by increasing $\chi^2$, with the relative epoch, gas density, temperature, freeze-out and H$_2$CO column density.}
\label{h2cofit}
\vspace{0.3 cm}
\begin{tabular}{|c|c|c|c|c|c|}
model & epoch & n(H$_{2}$) & T & FR & N(H$_2$CO)\\
     & (yr)   &(10$^{5}$ cm$^{-3}$) & (K) &\% & (cm$^{-2}$)\\ 
\hline
As1 & 1$\times10^3$ &  5      & 210 & 30 & 6.1$\times10^{14}$\\
B3  & 1$\times10^3$ &  5--10  & 110 & 60 & 6.5$\times10^{14}$\\
B11 & 4$\times10^3$ &  3--6   & 210 & 40 & 6.5$\times10^{14}$\\
B10 & 4$\times10^3$ &  5--10  & 210 & 20 & 5.1$\times10^{14}$\\
As1 & 2$\times10^3$ &  5      & 210 & 30 & 6.6$\times10^{14}$\\
B8  & 3$\times10^3$ &  2.5--5 & 210 & 40 & 6.0$\times10^{14}$\\
B10 & 3$\times10^3$ &  5--10  & 210 & 20 & 4.5$\times10^{14}$\\
\hline
\end{tabular}
\end{table}

\subsection{CH$_3$OH}

Methanol has always been considered a good tracer of the physical
condition of the emitting gas (e.g. \citealt{kale1997},
\citealt{leurini04}). In particular, it has been used as probe of
density in high-density medium (e.g. \citealt{menten1988}) and a
temperature estimate can be easily derived from the rotational
diagram, if multiple transitions are available. Moreover, the presence
of several transitions close in wavelength make it easy to observe
more lines simultaneously, increasing the number of observational
constraints and minimizing the relative calibration uncertainties.

The Boltzmann plot of the 8 methanol lines observed in the S1 clump
gives a column density N(CH$_3$OH)$\sim$ 10$^{16}$ cm$^{-2}$ and a
rotational temperature of only 16 K \citep{codella99}, well below
the 200 K we derived from the other molecules. However, in the
physical conditions typical of outflows the gas is not in LTE
condition and it is subthermally excited so that the rotational
temperature estimated from the Boltzmann plot is below the kinetic
temperature \citep{bachiller95}.

In the radiative transfer model the A and E types of methanol must be
considered as separate species since the transition between the two
symmetry states can not happen at a significant rate by radiative or
collisional processes. We assume the standard ratio of 1 between A and
E methanol as also suggested by the absence of a significant shift in
the A and E transitions in the Boltzmann plot. This choice has also
been verified a posteriori since we find that both A and E methanol
lines have similar deviation with respect to the observed lines.

From Table \ref{specie} we see that the 3$_{\rm K}$-2$_{\rm K}$ and 2$_{\rm K}$-1$_{\rm K}$ transitions are very close in frequency. The line rest frequencies of the CH$_3$OH 3$_0$-2$_0$ A$^+$ (145.103 GHz) and 3$_{-1}$-2$_{-1}$ E (145.097 GHz) transitions are separated by 12.4 \kms, those of the 3$_{-1}$-2$_{-1}$ E (145.097 GHz) and 3$_0$-2$_0$ E (145.094 GHz) by 6.2 \kms.
The 2$_0$-1$_0$A$^+$ (96.741 GHz) and 2$_{-1}$-1$_{-1}$ E (96.739 GHz) transitions are also separated by 6.2 \kms. Since the separation of the lines is similar or lower than the typical FWHM in the clump (10 \kms) we expect that 3$_{\rm K}$-2$_{\rm K}$ and 2$_{\rm K}$-1$_{\rm K}$ transitions overlap with each other. Indeed the lines at 145.1 and 96.7 GHz are blended; thus photons at these frequencies will see a higher optical depth than expected, because they are ``seeing'' the line profiles of neighbouring transitions. The SMMOL code does not yet have a line-overlap facility, so our modelling of these blended methanol transitions is semi-quantitative. For this reason we consider as an acceptable fit for the methanol all the models that produce line fluxes that differ less than 30\% with respect to the observed flux, instead of 10\% as used for the other species.

None of our models are able simultaneously to fit all of the 8
methanol transitions (5 lines of CH$_3$OH-E and 3 lines of
CH$_3$OH-A). In particular, we find that in models where the higher
energy transitions (5$_{\rm K}$-4$_{\rm K}$) are reproduced, the lower
energy transitions (3$_{\rm K}$-2$_{\rm K}$ and 2$_{\rm K}$-1$_{\rm
K}$) are largely underestimated. Conversely, models that reproduce the
lower energy transitions overestimate the transitions at higher
energy. This could be a direct consequence of the effects of line-overlap in the 3$_{\rm K}$-2$_{\rm K}$ and 2$_{\rm K}$-1$_{\rm K}$ transitions: the low energy transitions are simply ``seeing'' more optical depth than the higher energy transitions.

We did test, however, whether these inconsistencies are due to incorrectness
of our basic assumptions on the physical condition of the clump;
we computed further models where we assumed: {\it i)} a lower temperature of T=20 K or {\it ii)} a smaller size of the clump of 0.06 pc. Again, there is no single model able to fit both the lower and the higher energy transitions at the same time. Another possible explanation is that the high energy transitions trace a different component of the gas, possibly with different physical (and chemical) characteristics. Indeed the 5$_{\rm K}$-4$_{\rm K}$ lines seem to be
fitted with models that produce a methanol column density
N(CH$_3$OH)$\sim$5$\times10^{14}$ cm$^{-2}$ while the 3$_{\rm
K}$-2$_{\rm K}$ and 2$_{\rm K}$-1$_{\rm K}$ lines are fitted by models
that have a higher column density N(CH$_3$OH)$\sim$5$\times10^{15}$
cm$^{-2}$. Hence, we tried to separately fit the two components. For
the high energy transitions (the 5$_{\rm K}$-4$_{\rm K}$ lines) we
find that most of the models that fit the lines belong to Grid D (see
Table \ref{met241}), indicating that these transitions may trace a gas
compressed by the outflow (scenario D). Note that the models in Grid D
have a lower abundance of methanol than the models in the other
grids. This is due to reduced amount of frozen material, since
methanol is preferentially formed on the grains via hydrogenation of a
fraction of CO and H$_2$CO: in the scenario depicted in Grid D the
clump is formed by the compression of the outflow, and therefore the
timescale available for freeze out is quite short, as explained in
detail in Paper I. Moreover during the clump formation the temperature
also increases, slowing significantly the freeze out. While it is
possible that the high $J$ transitions trace a newly formed clump
(scenario of Grid D), it is also possible that this lower column
density component simply traces a smaller component of the clump. To
test this hypothesis we recomputed models A10 and As1 assuming a
smaller clump size of 0.04, 0.05 and 0.06 pc instead of 0.12 pc.
We found that both models A10 and As1 were able to fit the 3 CH$_3$OH lines of the 5$_{\rm K}$-4$_{\rm K}$ transition within 30\% with a clump size of 0.05 pc (see Table \ref{met241}). 

\begin{table}
\caption{List of the best models for the 3 lines of the CH$_3$OH 5$_{\rm K}$-4$_{\rm K}$ transition (both A and E type), ordered by increasing $\chi^2$, with the relative epoch, gas density, temperature, freeze-out and methanol column density. The last 2 models have a smaller clump size of 0.05 pc}
\label{met241}
\vspace{0.3 cm}
\begin{tabular}{|c|c|c|c|c|c|}
model & epoch & n(H$_{2}$) & T & FR & N(CH$_3$OH) \\
    & (yr)    &(10$^{5}$ cm$^{-3}$) & (K) &\% & (cm$^{-2}$)\\
\hline
D1  & (9-10)$\times10^4$ &  5--10 & 110 & 20& (4.2-4.8)$\times10^{14}$ \\
B12 & (8-9)$\times10^3$ &  2.5--5& 210 & 60& (4.3-4.7)$\times10^{14}$\\
D8  & 1$\times10^5$ &  5--10 & 100 & 20& 5.2$\times10^{14}$\\
D4  & 7$\times10^4$ &  3--5  & 100 & 20& 4.3$\times10^{14}$\\
\hline
A10 & (1-10)$\times10^4$ & 5 & 210 & 30 & (1.2-1.5)$\times10^{15}$ \\
As1 & (1-4)$\times10^3$  & 5 & 210 & 30 & 1.5$\times10^{15}$ \\
\hline
\end{tabular}
\end{table}

\begin{table}
\caption{List of the best models for the 5 lines of the CH$_3$OH3$_{\rm K}$-2$_{\rm K}$ and 2$_{\rm K}$-1$_{\rm K}$ transitions (both A and E type), ordered by increasing $\chi^2$, with the relative epoch, gas density, temperature, freeze-out and methanol column density.}
\label{met145}
\vspace{0.3 cm}
\begin{tabular}{|c|c|c|c|c|c|}
model & epoch & n(H$_{2}$) & T & FR & N(CH$_3$OH) \\
    & (yr)    &(10$^{5}$ cm$^{-3}$) & (K) &\% & (cm$^{-2}$)\\
\hline
B10 & 6$\times10^4$ & 5--10 & 210 & 20 & 7$\times10^{15}$\\
C3  & 7$\times10^4$ & 5--10 & 110 & 60 & 3$\times10^{15}$\\
\hline
\end{tabular}
\end{table}

\begin{figure}
\includegraphics[width=3.8cm,angle=90]{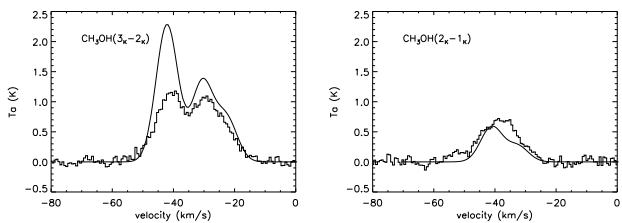}
\caption{The CH$_3$OH 3$_{\rm K}$-2$_{\rm K}$ and 2$_{\rm K}$-1$_{\rm K}$ lines observed in the clump S1 of the CB3 outflow (histogram) and the line profile predicted by the best fit model B10 at t=6$\times10^4$ yr (continuum line).}
\label{fig_met145}
\end{figure}

As discussed above the three 3$_{\rm K}$-2$_{\rm K}$ lines and the two 2$_{\rm K}$-1$_{\rm K}$ lines are closely blended due to line-overlap effects. Therefore to attempt a fit to the 5 CH$_3$OH
lines of the 3$_{\rm K}$-2$_{\rm K}$ and 2$_{\rm K}$-1$_{\rm K}$
transitions, we decided to calculate the $\chi ^2$ considering the sum
of the flux of all the blended lines rather than each line flux
separately. We find that two of our models fit the 5 lines (see Table
\ref{met145}). The best fit model is B10 (n(H$_2$)=(5-10)$\times10^5$
cm$^{-3}$, T$\sim$200 K, FR=20\%, Shock = Y) with a column density
N(CH$_3$OH)=7$\times10^{15}$ cm$^{-2}$. The best fit time is
t=6$\times10^4$ yr, although in B10 the methanol column density is
quite constant in time so it is not possible to constrain the age of
this component. In Fig. \ref{fig_met145} we show the best fit model:
although the total flux of the three blended 3$_{\rm K}$-2$_{\rm K}$
lines are still within the 30\% of tolerance, we do not achieve a good
fit for the CH$_3$OH 3$_0$-2$_0$ A$^+$ line at -41.5 \kms. We note
that in all the models the ratio between the 3$_0$-2$_0$ A$^+$ and the
3$_0$-2$_0$ E lines is always higher (up to a factor of 1.5) than the
observed value indicating a systematic trend to overestimate the
3$_0$-2$_0$ A$^+$ irrespective of the model parameters. This is due to the effect of line overlap that is not included in our radiative transfer model.

Despite the limitation due to the lack of line-overlap effect in the modeling, our analysis seems to indicate the presence of two different methanol components: one with lower column density N(CH$_3$OH) $\sim$ 5$\times10^{14}$-10$^{15}$ cm$^{-2}$, and possibly with smaller size, traced by the higher energy transitions, and another with higher column density N(CH$_3$OH) $\sim$ 7$\times$ 10$^{15}$ cm$^{-2}$ and probably at lower excitation conditions since it is traced by the lower energy lines. In our analysis we assumed the same temperature for both components; this may be unrealistic but we believe that in any case neither of the two components has a temperature less than 100 K, due to the presence of the outflow.
The possibility of different components or substructures within the clump was already put forward in Paper I for CB3 and partly confirmed by the better fits obtained for L1157 where the substructure is in fact observationally resolved. 

\section{DISCUSSION}

In the previous section we have shown that, if one considers each species 
separately, it is indeed possible to find a very good fit for each species. 
The column densities we find are always within a factor of 3 of
the ones derived by \citet{codella99} by using LTE or LVG models. 

We now attempt to find a model that simultaneously fits $all$ the
molecular transitions observed in the S1 clump. In order to take into
account the large uncertainties in the model for the global fit we
consider acceptable all the models where the line flux differs less
than 50\% from the observed flux. We choose this relatively high level
of tolerance because the line emission depends, directly or
indirectly, on many parameters that are not well known such as the
size of the emitting region, the density, the temperature and the
chemical abundances that in turn depend on other parameters such as
the percentage of depletion of the gas on to the grains, the efficiency
of some reactions, the density structure etc, and each
species/transition has a different sensitivity to each parameter. We
find that none of the 33 models is able to fit simultaneously all the
molecular lines within 50\% indicating, as also suggested in Paper I,
that indeed several gas components exist in the IRAM-30m beam.

\begin{figure}
\includegraphics[width=6cm,angle=90]{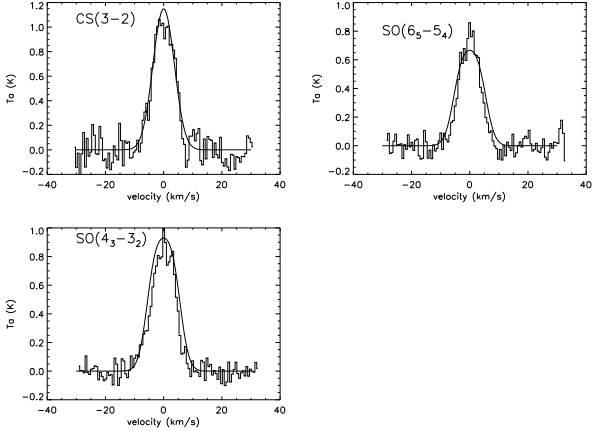}
\caption{The CS and SO lines observed in the clump S1 of the CB3 outflow (histogram) with the line profile predicted by the best fit model B9 at t=3$\times10^4$ yr (continuum line).}
\label{3linesfig}
\end{figure}

\begin{figure}
\includegraphics[width=5cm,angle=90]{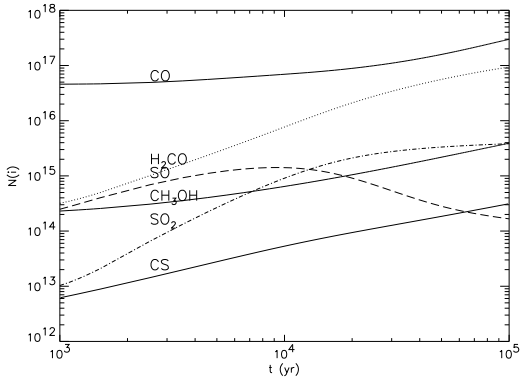}
\caption{Column densities versus time for the B9 model.}
\label{cdB9}
\end{figure}

\begin{figure}
\includegraphics[width=6cm,angle=90]{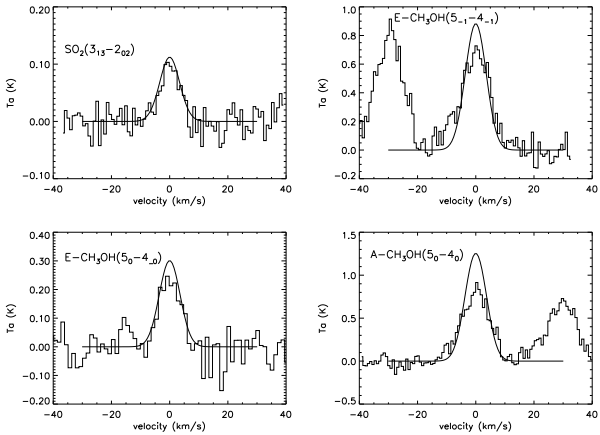}
\caption{The SO$_2$(3$_{13}$-2$_{02}$) and CH$_3$OH  (5$_{\rm K}$-4$_{\rm K}$) lines observed in the clump S1 of the CB3 outflow (histogram) with the line profile predicted by the best fit model As1 at t=4$\times10^4$ yr (continuum line). The size of the clump is d=0.05pc.}
\label{r025fig}
\end{figure}

In order to identify the different components we consider different
groups of molecules. Firstly, we consider all the S-bearing molecules
i.e. CS, SO, SO$_2$. None of the 33 models are able to fit all the 4
observed lines. In particular, while several models fit both the CS
and SO transitions (see Table \ref{sfit}) the same models do not fit
the SO$_2$ line that is usually fitted at the earlier times of
t$<$10$^{4}$ yr(see Table \ref{so2fit}). The best fit
model for CS and SO is the model B9 at a time of t=3$\times10^4$ yr
(see Fig. \ref{3linesfig}); however, at that time, the
SO$_2$(3$_{13}$-2$_{02}$) line is 8 times more intense than the
observed emission (see Fig. \ref{cdB9} where the column density of the
species is plotted versus time). In the same model the SO$_2$ line can
however be fitted at an early time, t=6$\times10^3$ yr, when the
column density is N(SO$_2$)=3$\times10^{14}$ cm$^{-2}$. Since in most
of the models the SO$_2$ abundance increases significantly with time,
the fact that the SO$_2$(3$_{13}$-2$_{02}$) line is fitted by young
epoch (see Table \ref{so2fit}) indicates that a low quantity of SO$_2$
is needed to fit the line. A low column density of SO$_2$ can be
obtained not just with a low chemical abundance but also with a
smaller size of the clump. In fact a smaller component in the clump has been already suggested by the analysis of the methanol lines. As we did for the methanol, we test this possibility considering the chemical models A10 and As1 with a smaller clump size of 0.04, 0.05 and 0.06 pc. In the A10 models the SO$_2$ abundance is always lower than that in the As1 model due to the lack of a high temperature phase in the former model. The A10 model does
not succeed in fitting the observed SO$_2$(3$_{13}$-2$_{02}$) line. On the
contrary, a very good fit can be obtained if we assume a size of 0.05
pc for the model As1 at the epoch t=4$\times10^4$ yr. At this epoch,
the model As1 also fits the 3 higher energy lines of methanol (see
Fig. \ref{r025fig}), as already shown in Section 4.5. The same model
does not fit the CS and SO lines. We therefore conclude that SO$_2$
and the higher energy methanol 5$_{\rm K}$-4$_{\rm K}$ lines may be
emitted from a smaller gas component with a size of d$\sim$0.05 pc,
n(H$_2$)$\sim 5\times10^5$ cm$^{-3}$ and T$\sim$200 K. This component
is associated with emission at high excitation of molecules produced
in high temperature gas thus it could trace the zone of stronger
interaction between the outflow and the clump.

\begin{table}
\caption{List of the best models for the SO(6$_5-5_4$), SO(4$_3-3_2$) and  CS(3-2) lines, ordered by increasing $\chi^2$, with the relative epoch, $\chi^2$, gas density temperature and freeze-out.}
\label{sfit}
\vspace{0.3 cm}
\begin{tabular}{|c|c|c|c|c|}
model & epoch & n(H$_{2}$) & T & FR \\
      & (yr) & (10$^{5}$ cm$^{-3}$) & (K) & \% \\
\hline
B9  & 3$\times$10$^4$ & 3--6  & 210 & 60\\
As1 & 1$\times$10$^3$ & 5     & 210 & 30\\
B4  & 2$\times$10$^3$ & 5--10 & 110 & 80\\
B12 & 2$\times$10$^4$ & 2.5--5& 210 & 60\\
B12 & 3$\times$10$^4$ & 2.5--5& 210 & 60\\
As1 & 2$\times$10$^3$ & 5     & 210 & 30\\
A1  & 2$\times$10$^3$ & 1     & 210 & 15\\
\hline
\end{tabular}
\end{table}

Next we attempt to fit simultaneously the CS and SO transitions
together with the 5 lower energy transitions of CH$_3$OH (3$_{\rm
K}$-2$_{\rm K}$ and 2$_{\rm K}$-1$_{\rm K}$). We find no models able
to fit simultaneously the 3 species within 50\%. In fact, from the
fitting of the single species (see Section 4.5) we see that the low energy methanol lines seem to trace a slightly higher density (model B10 with
n(H$_2$)=(5-10)$\times10^5$ cm$^{-3}$) than the CS and SO component
(model B9 with n(H$_2$)=(3-6)$\times10^5$ cm$^{-3}$), but the same temperature (T$\sim$200K) and very similar time (6$\times10^4$ yr and 3$\times10^4$ yr respectively). However the difference in the density is not so relevant, considering also the high uncertainties in the model especially for methanol because of line overlap effects, and we think that probably the low energy transitions of CH$_3$OH could be associated with the same extended gas component traced by CS and SO.

Finally we check whether the H$_2$CO 3$_{21}$-2$_{20}$ line could be associated with one of the two identified components. We find that the B9 model with clump size of 0.12 pc at t=3$\times10^4$ yr predicts a line about 25 times brighter than the observed one. Indeed, as said in Section 4.4, assuming a size of 0.12 pc the H$_2$CO can be fitted only at very early epoch (t$\le$4$\times10^3$ yr) while for more evolved times all the models predict far too much H$_2$CO. On the other hand, if we consider the smaller component, i.e. the As1 model at 4$\times10^4$ yr with clump size of 0.05 pc, the theoretical intensity is only 3 times higher than the observed one, while at the early epoch of 2 $\times10^4$ a very good fit (within 20\%) of the observed line is obtained. Although a global fit of the H$_2$CO with the SO$_2$ and high $J$ CH$_3$OH lines cannot be obtained within the 50\% of tolerance, we think that H$_2$CO does not trace the whole clump but is associated with the smaller gas component and it is probably emitted by a region even smaller than 0.05 pc.

The S1 clump has also been observed in the CO (2-1) line; however the
profile of this transition is very different with respect to all the
other lines: a blue non-Gaussian line wing and a self absorption at the
ambient LSR velocity (-38.5 km$^{-1}$) are present in the line. This
profile indicates that the line has a substantial contribution from
the large scale outflowing gas and also that part of the emission is
self absorbed by the ambient cloud. Moreover, \citet{codella99}
estimate a temperature of $\sim$10 K for the CO at all the line
velocities and they find that the (2-1) line is optically thick showing
that the emission is dominated by the outer, colder part of the
outflow rather than by the S1 clump. All these considerations lead us
to not consider the CO (2-1) line for the global fit of the S1
clump. However, we still ran the radiative transfer code for CO in
order to roughly evaluate if in the best fit models the contribution
of the clump to this line is negligible as we expected. Indeed, we
find that in the B9 model, that fits CS and SO, the predicted CO(2-1)
flux is only 23\% of the observed flux.

We find that some species are very sensitive to time, such as S-bearing molecules, in agreement with several previous studies (\citealt{hatchell98}, \citealt{viti01}, \citealt{wakelam2005}, \citealt{codella2005}), and they can therefore be used to determine the age of the S1 clump. We fit both the components of the clump with the same time $\sim$ 3--4$\times10^4$ yr
which is similar to the dynamical time of the S1 clump estimated from the SiO emission (t$\sim10^5$ yr, \citealt{codella99}). Note however that our starting point (t=0) in all scenarios is set to be when the outflow starts interacting with the surrounding environment hence the age we derive is not necessarily the age of the outflow.

\section{Conclusions}
A time-dependent chemical model has been coupled with a radiative
transfer code to model the observed molecular lines emission in the
small chemically rich clump S1 along the south lobe of the CB3
outflow. This proved to be a powerful tool in constraining the physical
and chemical  parameters of the gas, especially when there is more than one
transition of the same molecule.

We find that more than one model is able to fit the line emission of single species. Hence to constrain the physical parameters of our clump a multi-line analysis is needed. 

Different scenarios for the formation of the clump have been
investigated within a large range of physical conditions. Our results
show that most of the observed lines can not be fitted by models from
the scenarios A, C and D.  On the contrary, the models from scenario B
or As seem to reproduce most of our data. This is consistent with our
previous findings (see Paper I). We conclude that the observed dense 
clumps are, at least partially, formed prior to the advent of the outflow. 
The advent of the outflow on such clumps leads to a short phase of high
temperature (T$\sim$1000K) followed by fast cooling down to temperatures of 
$\sim$ 200 K.

In addition to the large scale outflowing gas traced by the non-Gaussian wings detected in the CO (2-1) line, we
identify two other components inside the large IRAM-30m beam. The
extended component (d=0.12 pc) is traced by CS and SO and probably by
the lower energy transitions (3$_{\rm K}$-2$_{\rm K}$ and 2$_{\rm
K}$-1$_{\rm K}$) of CH$_3$OH. This component is not uniform in density
but it has a density structure in the range 3$\times10^5$- 10$^6$
cm$^{-3}$. The second component is more compact, with size $\sim$0.05
pc and density $\sim$ 5$\times10^{5}$ cm$^{-3}$, and it is traced by
SO$_2$, H$_2$CO and the 5$_{\rm K}$-4$_{\rm K}$ CH$_3$OH lines. For both the
components we estimated a similar temperature (T$\sim$200 K) and time
(t$\sim$3$\times10^4$ yr) from when the clump has started to interact
with the outflow. Since the compact component is traced by
the higher energy transitions of CH$_3$OH, it is probably at higher
excitation and may trace the zone of stronger interaction between the outflow and the clump.  However, the analysis of the CH$_3$OH data needs to be confirmed by modelling of the data with a radiative transfer code that takes into account the effects of line-overlap (Gray \& Yates, in prep).

\section*{Acknowledgements}
We are grateful to Prof. D.R. Flower for providing us with the
molecular data of the CH$_3$OH-A. SV acknowledges individual financial support from a PPARC Advanced Fellowship. This work was partly carried out on the PSE supercomputer at the HiPerSPACE Computing Centre, UCL, which is funded by the UK Particle Physics and Astronomy Research Council.

\label{lastpage}

\end{document}